\documentclass[12pt,english]{article}
\usepackage[T1]{fontenc}
\usepackage[latin1]{inputenc}
\usepackage{graphics}
\usepackage{babel}
\begin{document}
\begin{center}
{\large Modified de Broglie-Bohm approach closer to classical
Hamilton-Jacobi  theory}
 
\vspace{.5 in}

Moncy V. John

{\it Department of Physics, St. Thomas College, Kozhencherri 689641,
 
Kerala, India.}

\vspace{.5in}

 {\bf Abstract}
\end{center}

A modified de Broglie-Bohm  (dBB) approach to quantum mechanics is
presented.  In this new deterministic theory, the problem of zero
velocity for  bound  states does not appear. Also this approach  is
equivalent to standard quantum mechanics when averages of dynamical
variables are taken, in  exactly the same way as in the original dBB
theory.  

\medskip
\noindent PACS No(s): 03.65.Bz

\newpage

In the de Broglie-Bohm  quantum theory of motion (dBB) \cite{dBB},
which  provides a deterministic theory of motion of quantum systems,
the following assumptions are made \cite{holland}:

(A1) An individual physical system comprises a wave propagating in
space and  time, together with a point particle which moves
continuously  
under the guidance of the wave. 

(A2) The wave is mathematically described by  $\Psi (x,t)$, a
solution to  the Schrodinger's wave  equation.

(A3) The particle motion is obtained as the solution $x(t)$ to the
equation  

\begin{equation}
\dot{x} = \frac {\hbar }{2im} \frac {\left( \Psi ^{\star }\frac
{\partial \Psi   }{\partial x}-\frac {\partial \Psi^{\star}}
{\partial x}\Psi \right)} {\Psi ^{\star} \Psi}, \label{eq:xdot1}
\end{equation}
with the initial condition $x=x(0)$. An ensemble of possible motions 
associated with the same wave is generated by varying $x(0)$. 

(A4) The probability that a particle in the ensemble lies between the
points  $x$ and $x+dx$ at time $t$ is given by $\Psi ^{\star }\Psi
\; dx$.

\medskip

In spite of its success as a deterministic theory, this scheme has
the drawback  that for many bound state problems of interest,  in
which the time-independent part $\psi$ of the wave function is real,
the velocity of the particle turns out to be  zero; i.e., the
particle is at rest whereas classically one would expect it to move
\cite{holland}.

To have a better understanding of this problem, let us recall that in
the    the classical Hamilton-Jacobi theory, the 
trajectory for a particle is obtained from the equation 

\begin{equation}
\frac {\partial S}{\partial t} + H\left( x,\frac {\partial
S}{\partial x}  \right) =0
\end{equation}
by first attempting to solve for the Hamilton-Jacobi function
$S=W-Et$ and then  integrating the equation of motion

\begin{equation}
p=m\dot{x} = \frac {\partial S}{\partial x} \label{eq:p}
\end{equation}
with respect to time. However, since the classical Hamilton-Jacobi
equation is  not adequate to describe the micro world, one resorts
to the Schrodinger equation  $i\hbar \partial \Psi/\partial t =
-(\hbar^{2}/2m)\partial^2 \Psi /\partial x^2 + V\Psi $ and solves for
the wave function $\Psi$. The basic idea of dBB is to put $\Psi$ in
the form $Re^{iS/\hbar}$ and then to identify $S$ as the
Hamilton-Jacobi function. The particle velocity is then given by Eq.
(\ref{eq:p}) above, and this leads to Eq. (\ref{eq:xdot1}). For  wave
functions whose space part is real, the Hamilton's characteristic
function  $W= constant$ and hence the velocity field is zero
everywhere, which is the problem mentioned above.

Another deterministic approach to quantum mechanics,  which also
claims the  absence of this problem,  is the trajectory
representation developed by Floyd \cite{floyd}. In one dimension,
Floyd uses the generalized Hamilton-Jacobi equation

\begin{equation}
\frac {(W^{\prime})^2}{2m} +V-E=-\frac{\hbar^2}{4m}\left[
\frac{W^{\prime  \prime \prime}}{W^{\prime}}-\frac{3}{2}\left( \frac
{W^{\prime \prime}}{W^{\prime}}\right)^2 \right],
\end{equation}
where  $W^{\prime}$
($^{\prime}$ denotes  $\frac{\partial }{\partial x}$) is the momentum
conjugate  to $x$.   Recently
Faraggi and Matone \cite{faraggi} have independently generated the
same  equation (referred to as the quantum stationary
Hamilton-Jacobi equation) from an equivalence principle free from any
axioms. The general solution to this nonlinear equation is given by
\cite{floyd}

\begin{equation}
W^{\prime} = \pm (2m)^{1/2}(a\phi^2 +b \theta^2 + c\phi \theta)^{-1}, 
\end{equation}
where 
($\phi$, $\theta $) is a set of normalized independent solutions of
the  associated  time-independent Schrodinger equation and ($a$,
$b$, $c$) is a set  of real coefficients such that $a,\; b>0$. Floyd
argues that the conjugate momentum $W^{\prime}$ is not the mechanical
momentum; instead, $m\dot{x} = m \partial E/\partial W^{\prime}$. The
equation of motion in the domain [$x,t$] is rendered by the Jacobi's
theorem. For stationarity, the equation of motion for the trajectory
time $t$, relative to its constant coordinate $\tau$, is given as a
function of $x$ by \cite{floyd,faraggi,carroll}

\begin{equation}
t-\tau = \frac {\partial W}{\partial E},
\end{equation}
where the trajectory is a function of ($a$, $b$, $c$) and $\tau $
specifies  the epoch. Each of these non unique trajectories manifest
a microstate of the Schrodinger wave function, even for  bound
states. Thus the Bohmian and trajectory representations differ 
significantly in the use of the equation of motion, though they are
based on equivalent generalized Hamilton-Jacobi equations. The
trajectory representation does not claim equivalence with standard
quantum mechanics in the predictions of  all observed phenomena nor 
does it explain the  concept of probability density inherent in the
Copenhagen interpretation.

In this Brief Report, we present a different and modified version of
the dBB   that surmounts the  problem mentioned in the introduction.
In addition,  it is closer to the classical Hamilton-Jacobi theory
than even the conventional dBB. We apply the scheme  to a few simple
problems and find that it is capable of providing a deeper insight
into the quantum phenomena. Also it is shown that like dBB,  the new
scheme is equivalent to standard quantum mechanics when averages of
dynamical variables are taken.

To this end, we first note that in the dBB, the
substitution $\Psi=Re^{iS/\hbar}$ in the Schrodinger equation  gives
rise to  two coupled partial differential equations, one of  which
contains an unknown quantum potential term, and this leads to a
situation apparently different from the classical Hamilton-Jacobi
theory for individual particles.  On the other hand, we  note
that   a change of variable $\Psi \equiv {\cal N}e^{iS/\hbar}$ in the
Schrodinger equation (where ${\cal N}$ is a constant) gives a
quantum-mechanical Hamilton-Jacobi equation \cite{goldstein}

\begin{equation}
\frac {\partial S}{\partial t} + \left[ \frac{1}{2m}\left( \frac
{\partial  S}{\partial x}\right)^2 +V\right]  = \frac{i\hbar}{2m}
\frac{\partial^2 S}{\partial x^2},
 \end{equation}
which  is closer to its classical
counterpart and which also has the correct classical limit.
This substitution     brings the classical  expression (\ref{eq:p})
for the  conjugate   momentum to  the form 

\begin{equation}
m\dot{x} = \frac {\hbar }{i} \frac {1}{\Psi}\frac {\partial
\Psi}{\partial x}.  \label{eq:xdot2}
  \end{equation}
We therefore attempt to modify the dBB scheme in a manner  similar to 
the latter approach; i.e., we retain assumptions A1, A2
and A4 whereas in A3, we identify $\Psi \equiv {\cal
N}e^{iS/\hbar}$ and use the    expression
(\ref{eq:xdot2}) as the equation of motion.

\medskip

To see how the scheme works, let us consider the example of the
ground state  solution $\Psi_0$ of the Schrodinger equation for the
harmonic oscillator in one dimension, the space part of which is 
real. We have

\begin{equation}
\Psi_0 ={\cal N}_0 e^{-\alpha^{2}x^{2}/2} e^{-iE_0t/\hbar}.
\end{equation}
The velocity field in the new scheme is given by Eq. (\ref{eq:xdot2})
as  

\begin{equation}
\dot{x}=  -\frac{\hbar}{im} \alpha^{2} x,\label{eq:xdotshm0}
 \end{equation}
whose solution is 

\begin{equation}
x=A e^{i\hbar \alpha^{2} t/m }.
\end{equation}

\begin{figure}[ht] 
\centering{\resizebox {0.8 \textwidth} {0.80 \textheight } 
{\includegraphics {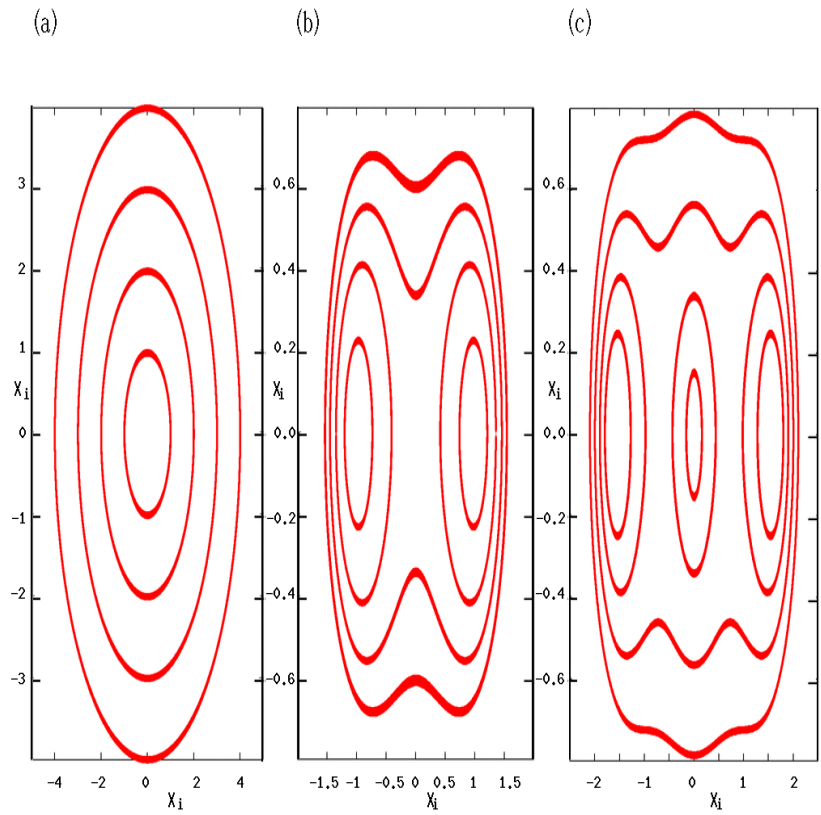}} 
\caption{Trajectories in the complex $X$-plane for  the harmonic
oscillator  ($X\equiv \alpha x$). (a) The $n=0$ case, where contours
are plotted for $X(0)$  =1, 2, 3 and 4. (b) The $n=1$ case, where
contours are plotted for $X(0)$ =1.2, 1.35, 1.45 and 1.55. (c) The
$n=2$ case, where contours are plotted for $X(0)$=1.8, 1.9, 2.0 and
2.1.}  \label{fig:shm}}     \end{figure}

This is an equation for a circle of radius $|A|=|x(0)|$ in the
complex plane  [Fig. \ref{fig:shm}(a)]. As usual in such mechanical
problems, we take the real part of this expression,

\begin{equation}
x_r=A \cos(\hbar \alpha^2t/m)
\end{equation}
[where it is chosen that at $t=0$,
$x(0)\equiv A$ is real], as the physical coordinate of the particle.
It shall  be noted that this is the same classical solution for a 
harmonic oscillator of frequency $\omega_0 = \hbar \alpha^2 /m$.

We can adopt this procedure to obtain the velocity field, also for
higher  values of the quantum number n. For $n=1$, we have

\begin{equation}
\Psi_1 = {\cal N}_1 e^{-\alpha^{2}x^{2}/2}\; 2\alpha  x \; e^{-iE_1
t/\hbar},  \end{equation}
from which

\begin{equation}
\dot{x}=\frac{\hbar}{im} \left(-\alpha^{2} x +
\frac{1}{x}\right).\label{eq:xdotshm1}  
\end{equation}
The solution to this equation can be written as 

\begin{equation}
(\alpha x -1)(\alpha x +1)=A e^{2i\hbar\alpha ^{2} t/m}
\end{equation}
or

\begin{equation}
x=\frac {1}{\alpha }\sqrt {1+A e^{2i\hbar\alpha ^{2} t/m}}.
\end{equation}
Here, the solution is a product of two circles centered about $\alpha
x=\pm  1$, which is plotted in Fig. \ref{fig:shm}(b). The physical
coordinate of the particle  is again given by  the real part of this
expression. For $n=2$, the solution can similarly be constructed as 

\begin{equation}
4\alpha x \left( \alpha x -\sqrt{5/2}\right)^{2}\left(\alpha x
+\sqrt{5/2}\right)^{2}=A e^{5i\hbar\alpha  ^{2} t/m}.
 \end{equation}
The trajectory in the complex plane is plotted in Fig.
\ref{fig:shm}(c).  Note that in all these cases, the velocity fields
are not zero everywhere.

Now let us apply the procedure to some other stationary states which
have  complex wave functions. For a plane wave, we have

\begin{equation}
\Psi =A e^{ikx}e^{-iEt/\hbar},
\end{equation}
so that $\dot{x}=\hbar k/m$ and this gives

\begin{equation}
x=x(0)+\frac{\hbar k}{m} t,
\end{equation}
the classical solution for a free particle.

\begin{figure}[ht] 
\centering{\resizebox {0.8 \textwidth} {0.8 \textheight }  
{\includegraphics {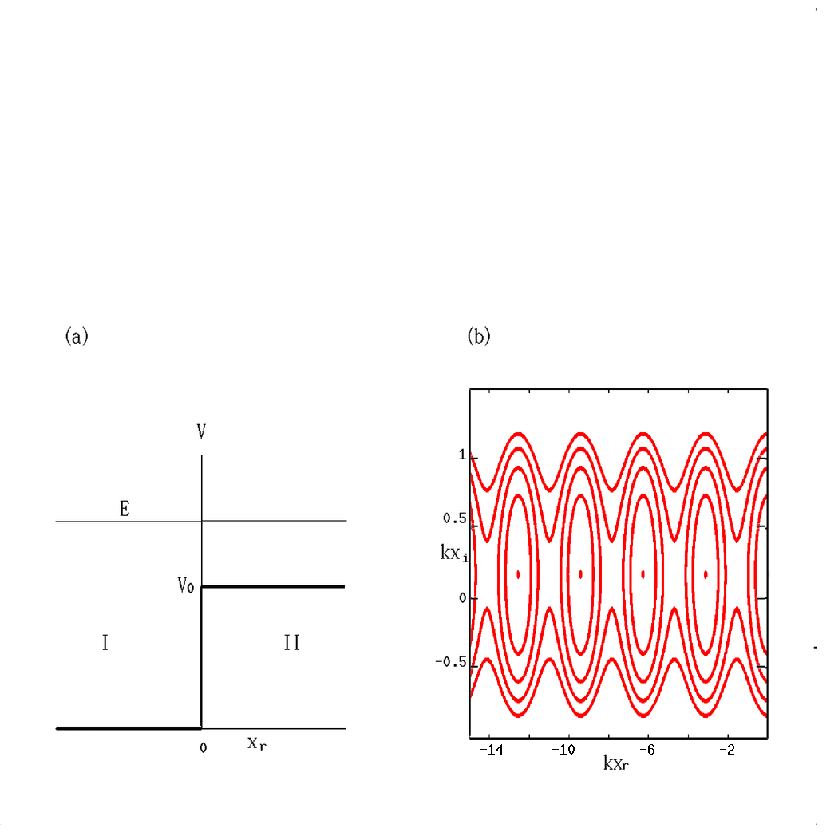}} 
\caption{(a) Potential step, with $V=0$ for  $x_r<0$, $V$ = $V_0$ for 
$x_r >0$ and energy $E>V_0$. (b) Trajectories in the complex plane for 
particles approaching the potential step. Contours for $c$ = -4, -3,
-2, -1  and 0 are plotted. }  \label{fig:step}}    \end{figure}

As another example, consider a particle with energy $E$ approaching a
potential step with height $V_0$, shown in Fig. \ref{fig:step}(a). In 
region I, we have

\begin{equation}
\psi_I = e^{ikx}+Re^{-ikx}, \qquad k=\sqrt{2mE/\hbar^2}
\end{equation}
and in region II,

\begin{equation}
\psi_{II} = Te^{iqx}, \qquad q=\sqrt{2m(E-V_0 )/\hbar^2}.
\end{equation}
 The velocity fields in the two regions are given by

\begin{equation}
\dot{x}_I = \frac {\hbar k}{m}\left(\frac{e^{ikx}-Re^{-ikx}
}{e^{ikx}+Re^{-ikx}}\right)
\end{equation}
and 

\begin{equation}
\dot{x}_{II} =\frac{\hbar q}{m},
\end{equation}
respectively. The contours in the complex $x$-plane in region I, for
a  typical value of the reflection coefficient $r\equiv R^2=1/2$,
are given by

\begin{equation}
\sqrt{2}  \cos 2kx_{Ir}-e^{2kx_{Ii}}-\frac{1}{2}e^{2kx_{Ii}}=c
\end{equation}
[where $x_{Ir}$ and $x_{Ii}$ are, respectively, the real and
imaginary parts of  $x_I$], and are plotted in Fig.
\ref{fig:step}(b). Note that also this case  is significantly
different from the corresponding dBB solution.

Lastly, we consider a nonstationary wave function, which is a
spreading wave  packet. Here, let the propagation constant $k$ has a
Gaussian spectrum with a width $\Delta k \sim 1/\sigma $ about a mean
value $\bar{k}$. The wave function is given by 

\begin{equation}
\Psi (x,t)={\cal N} \left( \frac {2\pi \sigma }{\sigma^2 +i\hbar 
t/m}\right)^{1/2} \exp \left[ - \frac {(x-i\sigma
^2\bar{k})^2}{2(\sigma^2 +  i\hbar t/m)} - \frac { \sigma^2 +
\bar{k}^2}{2} \right]. \end{equation}
The velocity field is obtained from (\ref{eq:xdot2}) as 

\begin{equation}
\dot{x} = - \frac {\hbar}{im}  \left( \frac {x-i\sigma ^2
\bar{k}}{\sigma^2+i\hbar t/m}\right) 
\end{equation}
Integrating this expression, we get

\begin{equation}
x(t)= x(0)+ \frac{\hbar \bar{k}}{m} t+ i \frac{\hbar}{m}
\frac{x(0)}{\sigma^2}t. 
 \end{equation}
Separating the real and imaginary parts of this equation (and
assuming  $\bar{k}$ is real), we get,

\begin{equation}
x_r(t)= x_r (0)+ \frac{\hbar \bar{k}}{m} t+
 \frac{\hbar}{m}   \frac{x_i (0)}{\sigma^2}t 
\end{equation}
and 

\begin{equation}
x_i (t)= x_i (0)+ 
\frac{\hbar}{m}   \frac{x_r (0)}{\sigma^2}t .
\end{equation}
For the particle with $x(0)=0$, we obtain $x_r(t) =(\hbar
\bar{k}/m)t$  and $x_i(t)=0$; i.e., this particle remains at the
center of the wave packet. Other particles assume different positions
$x(t)$ at time $t$ as given by the above expression, which indicates
the spread of the wave packet.

\medskip

A new element in the present approach to quantum mechanics (though it
is quite  familiar in elementary mechanical problems) is that  the
particle coordinates are having real and imaginary parts. Similar is
the case for the velocity, given by Eq. (\ref{eq:xdot2}). If we are
interested in the real part of this velocity, then one can write

\begin{equation}
\dot{x}_r = \frac {\hbar }{2im} \frac {\left[ \Psi ^{\star }\frac
{\partial  \Psi  }{\partial x}-(\frac {\partial \Psi} {\partial
x})^{\star}\Psi \right]} {\Psi ^{\star} \Psi}, \label{eq:xdot4}
\end{equation}
an expression quite analogous to the expression for velocity field in
the dBB  scheme (though not identical with it). One can also write
$\dot{x}_r$ as

\begin{equation}
\dot{x}_r = \frac {\hbar }{2im} \frac {\left[ \Psi ^{\star }\frac
{\partial  \Psi  }{\partial x_r}-(\frac {\partial \Psi} {\partial
x_r})^{\star}\Psi \right]} {\Psi ^{\star} \Psi}, \label{eq:xdot5}
\end{equation}
where use is made of the fact that for complex derivatives, $\frac
{\partial  \Psi  }{\partial x}=\frac {\partial \Psi  }{\partial
x_r}$. This equation is identical to the dBB velocity field at all
points $x=x_r$.

\medskip

Now consider an ensemble of particles, whose initial density function
along  $x=x_r$ is given by 

\begin{equation}
P(x_r, t=0)=\Psi^{\star}(x=x_r, t=0)\Psi (x=x_r, t=0).
\end{equation}
Now let us ask whether $P(x_r, t)=\Psi^{\star}(x=x_r, t)\Psi (x=x_r,
t)$ be  identified as the density function along $x=x_r$ at any time
$t$, given the velocity field (\ref{eq:xdot2}). For this to be in the
affirmative, the conservation equation

\begin{equation}
\frac{\partial P}{\partial t}+\frac {\partial}{\partial x_r}(P\dot
{x}_r) =0  \end{equation}
must hold, with $\dot{x}_r$ given by Eq. (\ref{eq:xdot4}) above. It
is easy to  see that this is  true, provided $V(x=x_r)$ is
real. Thus one can retain the assumption (A4), with
$\Psi^{\star}(x=x_r, t)\Psi  (x=x_r, t)$ as the density function  at
any time $t$. This guarantees that the averages of dynamical
variables $O$ constructed using the real variables $x_r$ and $t$,
computed over the measure $P$ will necessarily agree with the quantum
mechanical expectation values of the corresponding hermitian
operators $\hat {O}$ at all future times $t$; i.e., 

\begin{equation}
\hat{O} = \int_{t=constant} PO dx_r=\int \Psi^{\star}(x=x_r, t) O
\Psi (x=x_r,  t)dx_r,
\end{equation}
in the same way as in dBB. Thus the new scheme is equivalent to
standard  quantum mechanics when  averages of dynamical variables
are taken, just as in the case of the original dBB approach.

\medskip

Generalization of the formalism to more than one dimension is not
attempted  in this Brief Report. Its application to those other
physical problems of interest and a host of other issues, which need
careful analysis, shall be addressed in future publications.

Finally, we note that the dBB identification $\Psi =Re^{iS/\hbar}$
does not   help to utilize  all the information contained  in the
wave function while solving   the equation of motion
$m\dot{x}=\partial S/\partial x $, though it provides a conservation
equation for $R^2$.  In the present formulation, we obtain a quantum
Hamilton-Jacobi equation, which is closer to the classical one and
which does not have   any exotic quantum potential term. But still we
could  get the conservation equation for probability density, as
demonstrated above. The positive results we obtained for the harmonic
oscillator and potential step problems themselves are indicative of
the deep insight obtainable in such problems by the use of the new
scheme. The price we had to pay for this is the appearance of an
imaginary part in the position and velocity of the particle, but  the
technique is considered to be quite normal in elementary mechanical
problems (and not yet in quantum mechanical ones). Therefore, it is 
desirable to  further explore  the consequences of the new scheme.

\begin{center}
{\bf Acknowledgements}
\end{center}

It is a pleasure to thank  Professor K. Babu Joseph, Sajith and
Satheesh  for enlightening discussions.

\end{document}